\documentclass[aps,preprint,showpacs]{revtex4}
\usepackage{psfig}
\begin{document}           
\draft
\title{Possible Negative Pressure States in the Evolution of the Universe}
\author{A. Kwang-Hua Chu} 
\affiliation{P.O. Box 30-15, Shanghai 200030, PR China}
\begin{abstract}
Hydrodynamic derivation of the
entrainment of matter induced by a surface elastic wave
propagating along the flexible vacuum-matter interface is
conducted by considering the nonlinear coupling between the
interface and the rarefaction effect.  The critical reflux values associated with the
product of the second-order (unit) body forcing and the Reynolds
number (representing the viscous dissipations) decrease as the Knudsen number
(representing the rarefaction measure) increases from zero to 0.1.
We obtained the critical bounds for matter-freezed or zero-volume-flow-rate states
corresponding to specific Reynolds numbers
(ratio of wave inertia and viscous dissipation effects)
and wave numbers which might be linked to
the dissipative evolution of the Universe. Our results also show that for certain
time-averaged evolution
of the matter (gas) there might be existence of negative pressure.
\newline

\end{abstract}
\maketitle                 
\bibliographystyle{plain}
\section{Introduction}
The mean cosmic density of dark matter (plus baryons) is now pinned down to
be only ca. 30\% of the so-called critical density corresponding
to a 'flat'-Universe.
However, other recent evidence|microwave background anisotropies,
complemented
by data on distant supernovae|reveals that our Universe actually is 'flat', but that
its dominant ingredient (ca. 70\% of the total mass energy) is something
quite unexpected:
'dark energy' pervading all space, with negative pressure.
We do know that this material is very dark and that
it dominates the internal kinematics, clustering properties and motions
of galactic
systems. Dark matter is commonly associated to weakly interacting particles (WIMPs), and
can be described as a fluid with vanishing pressure.
It plays a crucial role in the formation and evolution of structure
in the
universe and it is unlikely that galaxies could have formed without
its presence [1-2].
Analysis of cosmological mixed dark matter models in spatially flat
Friedmann
Universe with zero $\Lambda$ term have been presented before.
For example, we can start from the Einstein action describing the
gravitational forces in the presence of the cosmological constant [3]
\begin{displaymath}
 S=\frac{1}{2 k_N^2} \int \sqrt{-g}\, R - \int \sqrt{-g}\, \Lambda + S_m,
\end{displaymath}
where $S_m$ is the contribution of the matter and radiation, $k_N^2=8\pi G_N=8\pi/M_P^2$,
$G_N$ is the Newton's constant and $M_P = 1.22 \times 10^{19} $GeV/$c^2$ is
the Planck mass. The set of equations governing the evolution of the
universe is completed by the Friedman equations for
the scale factor ($a(t)$)
\begin{displaymath}
 (\frac{\dot{a}}{a})^2=\frac{8\pi G_N}{3}(\rho+\Lambda)-\frac{k}{a^2},
\end{displaymath}
\begin{displaymath}
 (\frac{\ddot{a}}{a})^2=-\frac{4\pi G_N}{3}(\rho+\Lambda+3(p-\Lambda)),
\end{displaymath}
where $k=0$ is for the flat-Universe, $\rho$ and $p$ is the density
and pressure [4-5]. A large majority of dark energy models describes
dark energy in terms of the equation of state (EOS)
$p_d = \omega\,\rho_d$,
where $\omega$ is the parameter of the EOS, while $p_d$
and $\rho_d$ denote the pressure and the energy density
of dark energy, respectively. The value $\omega = -1$ is
characteristic of the cosmological constant, while the
dynamical models of dark energy generally have $\omega \ge
-1$. The case of the growing
cosmological term $\Lambda$ and its implications for the asymptotic
expansion of the universe and the destiny of
the bound systems have been studied in [5] using above system of equations.
Their results showed that even for very slow
growth of $\Lambda$ (which satisfies all the conditions on the
variation of $G_N$),
in the distant future the gravitationally bound systems
become unbound, while the nongravitationally bound
systems remain bound. \newline
Meanwhile, it is convenient to express the mean densities $\rho_i$ of various
quantities in the Universe
in terms of their fractions relative to the critical density:
$\Omega_i=\rho_i /\rho_{crit}$. The theory
of cosmological inflation strongly suggests that the total density should
be very close
to the critical one ($\Omega_{tot} \sim 1$), and this is supported by
the available data on the
cosmic microwave background (CMB) radiation. The fluctuations
observed in the CMB at a level ca. 10$^{-5}$ in amplitude exhibit a peak
at a partial
wave $l \sim 200$, as would be produced by acoustic oscillations in
a flat Universe with
$\Omega_{tot} \sim 1$. At lower partial waves, $l \gg 200$,
the CMB fluctuations are believed to be
dominated by the Sachs-Wolfe effect due to the gravitational potential,
and more
acoustic oscillations are expected at $l > 200$,
whose relative heights depend on the
baryon density $\Omega_b$. At even larger values, $l \ge 1000$,
these oscillations should be
progressively damped away [1-2,3,6]. \newline
Influential only over the largest of scales-the cosmological
horizon-is the outermost species of invisible
matter: the vacuum energy (also known by such names
as dark energy, quintessence, $x$-matter, the zero-point
field, and the cosmological constant $\Lambda$).
If there is no exchange of energy between vacuum and
matter components, the requirement of general covariance implies the time
dependence of the gravitational constant $G$. Thus, it is interesting to
look at the interacting behavior between the vacuum (energy) and the matter
from the macroscopic point of view. One related issue, say, is
about the dissipative matter of the flat Universe
immersed in vacua [7] and the other one is the macroscopic Casimir effect with
the deformed boundaries [8].
\newline
%
Theoretical (using the Boltzmann equation) and experimental
studies of interphase nonlocal transport phenomena which appear as
a result of a different type of
nonequilibrium representing propagation of a surface elastic wave
have been performed since late 1980s [9-10].
These are relevant to rarefied gases (RG) flowing along deformable elastic
slabs with the dominated
parameter being the Knudsen number (Kn = mean-free-path/$L_d$,
mean-free-path (mfp) is the mean free path of the gas, $L_d$ is proportional to
the
distance between two slabs) [11-13]. The role of the Knudsen number is
similar to that of
the Navier slip parameter $N_s$ [14]; here, $N_s = \mu S/d$ is the dimensionless
Navier slip parameter; S is a proportionality constant as $u_s = S \tau$,
$\tau$ :
the shear stress of the bulk velocity; $u_s$ : the dimensional slip velocity;
for a
no-slip case, $S = 0$, but for a no-stress condition. $S=\infty$, $\mu$ is the
fluid viscosity, $d$ is one half of the distance between upper and
lower slabs). \newline
Here, the transport driven by the wavy elastic vacuum-matter interface will
be presented. The flat-Universe is presumed and the corresponding matter is immersed
in vacua with the interface being flat-plane like.
We adopt the macroscopic or hydrodynamical approach and simplify the original
system of equations (related to the momentum and mass transport)
to one single higher-order quasi-linear partial differential
equation in terms of the unknown stream function. In this study,
we shall assume that the Mach number $Ma \ll 1$, and the governing
equations are the incompressible Navier-Stokes equations which are
associated with the relaxed slip velocity boundary conditions
along the interfaces [11-14]. We then introduce the perturbation
technique so that we can solve the related boundary value problem
approximately. To consider the originally quiescent gas for
simplicity, due to the difficulty in solving a fourth-order
quasi-linear complex ordinary differential equation (when the wavy
boundary condition are imposed), we can finally get an
analytically perturbed solution and calculate those physical
quantities we have interests, like, time-averaged transport or entrainment,
perturbed velocity functions, critical unit body forcing corresponding to
the freezed or zero-volume-flow-rate states. These results might be closely linked to
the vacuum-matter interactions (say, macroscopic Casimir effects) and
the evolution of the Universe (as mentioned above : the critical density [1-2]).
Our results also show that for certain
time-averaged evolution
of the matter (the maximum speed
of the matter (gas) appears at the center-line) there might be existence of
negative-pressure states.
\section{Formulations}
We consider a two-dimensional matter-region of uniform thickness
which is approximated
by a homogeneous rarefied gas (Newtonian viscous fluid). The equation of motion is
\begin{equation}
  (\lambda_L+\mu) \mbox{grad}\, \mbox{div}\, {\bf u} +\mu \nabla {\bf u} +\rho {\bf p}=
  \rho \frac{\partial^2 {\bf u}}{\partial t^2},
\end{equation}
where $\lambda_L$ and $\mu$ are Lam\'{e} constants, ${\bf u}$ is the
displacement field (vector),
$\rho$ is the mass density and ${\bf p}$ is the body force for unit mass.
The Navier-Stokes equations, valid for Newtonian fluids
(both gases and liquids), has been a mixture of continuum fluid
mechanics ever since 1845 following the
seemingly definitive work of Stokes  and others [15], who proposed the
following
rheological constitutive expression for the fluid deviatoric or viscous
stress (tensor) ${\bf T}$ :
${\bf T}= 2\mu \nabla {\bf u}+\lambda_L {\bf I} \nabla \cdot {\bf u}$. \newline
The flat-plane
boundaries of this matter-region  or the vacuum-matter interfaces
are rather flexible and presumed to be elastic, on which are imposed
traveling sinusoidal waves of small amplitude $a$ (possibly due to vacuum fluctuations).
The vertical displacements of the upper and lower interfaces ($y=h$ and
$-h$) are thus presumed to be $\eta$ and $-\eta$, respectively,
where $\eta=a \cos [2\pi (x-ct)/\lambda$], $\lambda$ is the
wave length, and $c$ the wave speed. $x$ and $y$ are Cartesian
coordinates, with $x$ measured in the direction of wave
propagation and $y$ measured in the direction normal to the mean
position of the vacuum-matter interfaces.
It would be expedient to simplify these equations by
introducing dimensionless variables. We have a characteristic
velocity $c$ and three characteristic lengths $a$, $\lambda$, and
$h$. The following variables based on $c$ and $h$ could thus be
introduced :
\begin{displaymath}
 x'=\frac{x}{h}, \hspace*{3mm} y'=\frac{y}{h}, \hspace*{6mm}
 u'=\frac{u}{c}, \hspace*{3mm} v'=\frac{v}{c}, \hspace*{2mm}
 \eta'=\frac{\eta}{h}, \hspace*{3mm} \psi'=\frac{\psi}{c\,h}, \hspace*{4mm}
 t'=\frac{c\,t}{h}, \hspace*{3mm} p'=\frac{p}{\rho c^2},
\end{displaymath}
where $\psi$ is the dimensional stream function, $u$ and $v$ are the velocities
along the $x$- and $y$-directions; $\rho$ is the density, $p$ (its gradient)
is related to the (unit) body
forcing. The primes could be dropped in the following. The amplitude
ratio $\epsilon$, the wave number $\alpha$, and the Reynolds
number (ratio of wave inertia and viscous dissipation effects) $Re$ are defined by
\begin{displaymath}
 \epsilon=\frac{a}{h}, \hspace*{4mm} \alpha=\frac{2 \pi h}{\lambda},
 \hspace*{4mm} Re =\frac{c\,h}{\nu}.
\end{displaymath}
We shall seek a solution in the form of a series in the parameter
$\epsilon$ :
\begin{displaymath}
 \psi=\psi_0 +\epsilon \psi_1 + \epsilon^2 \psi_2 + \cdots,
\end{displaymath}
\begin{displaymath}
\frac{\partial p}{\partial x}=(\frac{\partial p}{\partial
 x})_0+\epsilon (\frac{\partial p}{\partial x})_1 +\epsilon^2 (\frac{\partial
 p}{\partial x})_2 +\cdots,
\end{displaymath}
with $u=\partial \psi/\partial y$, $v=-\partial \psi/\partial x$.
The 2D (x- and y-) momentum equations and the equation of
continuity could be in terms of the stream function $\psi$ if the
$p$-term (the specific body force density, assumed
to be conservative and hence expressed as the gradient of a time-independent
potential energy function) is eliminated. The final governing equation is
\begin{equation}
 \frac{\partial}{\partial t} \nabla^2 \psi + \psi_y \nabla^2 \psi_x
 -\psi_x \nabla^2 \psi_y =\frac{1}{Re}\nabla^4 \psi,
\hspace*{12mm} \nabla^2 \equiv\frac{\partial^2}{\partial x^2}
+\frac{\partial^2}{\partial y^2}   ,
\end{equation}
and subscripts indicate the partial differentiation. Thus, we have
\begin{equation}
 \frac{\partial}{\partial t} \nabla^2 \psi_0 +\psi_{0y} \nabla^2
 \psi_{0x}-\psi_{0x} \nabla^2 \psi_{0y}=\frac{1}{Re} \nabla^4 \psi_0
 ,
\end{equation}
\begin{equation}
 \frac{\partial}{\partial t} \nabla^2 \psi_1 +\psi_{0y} \nabla^2
 \psi_{1x}+\psi_{1y}\nabla^2 \psi_{0x}-\psi_{0x} \nabla^2 \psi_{1y}-
 \psi_{1x} \nabla^2 \psi_{0y}=\frac{1}{Re} \nabla^4 \psi_1
 ,
\end{equation}
\begin{displaymath}
 \frac{\partial}{\partial t} \nabla^2 \psi_2 +\psi_{0y} \nabla^2
 \psi_{2x}+\psi_{1y}\nabla^2 \psi_{1x}+\psi_{2y} \nabla^2 \psi_{0x}-
\end{displaymath}
\begin{equation}
 \hspace*{25mm} \psi_{0x} \nabla^2 \psi_{2y}- \psi_{1x} \nabla^2 \psi_{1y}-
 \psi_{2x}\nabla^2 \psi_{0y}=\frac{1}{Re} \nabla^4 \psi_2
 ,
\end{equation}
and other higher order terms. The (matter) gas is subjected to boundary
conditions imposed by the symmetric motion of the vacuum-matter interfaces and the
non-zero slip velocity : $u=\mp$ Kn $\,du/dy$ [11-13], $v=\pm
\partial \eta/\partial t$ at $y=\pm (1+ \eta)$, here Kn=mfp$/(2h)$.
The boundary conditions may be expanded in powers of $\eta$ and
then $\epsilon$ :
\begin{displaymath}
 \psi_{0y}|_1 +\epsilon [\cos \alpha (x-t) \psi_{0yy}|_1 +\psi_{1y}|_1
 ]+\epsilon^2 [\frac{\psi_{0yyy}|_1}{2} \cos^2 \alpha (x-t)
 +\psi_{2y}|_1 +
\end{displaymath}
\begin{displaymath}
 \hspace*{3mm}\cos \alpha (x-t) \psi_{1yy}|_1 ] +\cdots =-\mbox{Kn} \{\psi_{0yy}|_1
 +\epsilon [\cos \alpha (x-t) \psi_{0yyy}|_1 +\psi_{1yy}|_1 ]+
\end{displaymath}
\begin{equation}
 \hspace*{3mm}\epsilon^2 [\frac{\psi_{0yyyy}|_1}{2} \cos^2 \alpha (x-t)+
 \cos \alpha (x-t) \psi_{1yyy}|_1 +\psi_{2yy}|_1 ] +\cdots \}
 ,
\end{equation}
\begin{displaymath}
 \psi_{0x}|_1 +\epsilon [\cos \alpha (x-t) \psi_{0xy}|_1 +\psi_{1x}|_1
 ]+\epsilon^2 [\frac{\psi_{0xyy}|_1}{2} \cos^2 \alpha (x-t)+
\end{displaymath}
\begin{equation}
 \hspace*{3mm} \cos \alpha (x-t) \psi_{1xy}|_1 + \psi_{2x}|_1 ] +\cdots =
  -\epsilon \alpha \sin \alpha (x-t).  
\end{equation}
Equations above, together with the condition of symmetry and a
uniform  $(\partial
p/\partial x)_0$, yield :
\begin{equation}
 \psi_0 =K_0 [ (1+2 \mbox{Kn}) y-\frac{y^3}{3}],  \hspace*{24mm}
 K_0=\frac{Re}{2}(-\frac{\partial p}{\partial x})_0 , 
\end{equation}
\begin{equation}
 \psi_1 =\frac{1}{2}\{\phi(y) e^{i \alpha (x-t)}+\phi^* (y) e^{-i \alpha
 (x-t)} \} , 
\end{equation}
where the asterisk denotes the complex conjugate. A substitution
of $\psi_1$ into Eqn. (3) yields
\begin{displaymath}
 \{\frac{d^2}{d y^2} -\alpha^2 +i \alpha Re [1-K_0 (1-y^2+2 \mbox{Kn})]\}
 (\frac{d^2}{d y^2} -\alpha^2) \phi -2 i\alpha K_0 Re \,\phi =0  
\end{displaymath}
or
if originally the (matter) gas is quiescent : $K_0 = 0$ (this
corresponds to a free (vacuum) pumping case)
\begin{equation}
 (\frac{d^2}{d y^2} -\alpha^2) (\frac{d^2}{d y^2} -\bar{\alpha}^2) \phi
 =0 ,   \hspace*{24mm} \bar{\alpha}^2= \alpha^2 -i \alpha Re. 
\end{equation}
The boundary conditions are
\begin{equation}
 \phi_y (\pm 1) \pm\phi_{yy} (\pm 1) \mbox{Kn}=2 K_0 (1\pm \mbox{Kn})=0, \hspace*{24mm}
 \phi (\pm 1)=\pm 1 .  
\end{equation}
Similarly, with
\begin{equation}
  \psi_2=\frac{1}{2} \{D(y)+E(y) e^{i 2\alpha (x-t)} +E^* (y) e^{-i
  2\alpha (x-t)} \} ,  
\end{equation}
we have
\begin{equation}
 D_{yyyy}=-\frac{i \alpha Re}{2} (\phi \phi^*_{yy}-\phi^*
 \phi_{yy})_y , 
\end{equation}
\begin{displaymath}
 [\frac{d^2}{d y^2} -(4\alpha^2 -2 i \alpha Re) ] (\frac{d^2}{d y^2} -4\alpha^2)
  E-i 2 \alpha Re K_0 (1-y^2+2 \mbox{Kn})
\end{displaymath}
\begin{equation}
 \hspace*{12mm}(\frac{d^2}{d y^2} -4\alpha^2) E + i 4 \alpha K_0 Re E +
 \frac{i \alpha Re}{2}(\phi_y \phi_{yy}-\phi \phi_{yyy}) =0 ;  
\end{equation}
and the boundary conditions
\begin{equation}
 D_y (\pm 1) +\frac{1}{2} [\phi_{yy} (\pm 1)+\phi^*_{yy} (\pm 1)] -2
 K_0= \mp \mbox{Kn}  \{\frac{1}{2} [\phi_{yyy} (\pm 1) +
\phi^*_{yyy} (\pm 1)]+ D_{yy} (\pm 1) \},  
\end{equation}
\begin{equation}
 E_y (\pm 1)+\frac{1}{2} \phi_{yy} (\pm 1) -\frac{K_0}{2} =\mp \mbox{Kn} [\frac{1}{2}
 \phi_{yyy} (\pm 1) + E_{yy} (\pm 1) ] ,
\end{equation}
\begin{equation}
 \hspace*{24mm} E(\pm 1)+\frac{1}{4} \phi_y  (\pm 1) =0   
\end{equation}
where $K_0$ is zero in Eqns. (13-16).
After lengthy algebraic manipulations, we obtain
\begin{displaymath}
 \phi=c_0 e^{\alpha y}+c_1 e^{-\alpha y}+c_2 e^{\bar{\alpha} y}+
      c_3 e^{-\bar{\alpha} y} ,
\end{displaymath}
where $c_0=(A+A_0)/Det$, $c_1=-(B+B_0)/Det$, $c_2=(C+C_0)/Det$,
$c_3=-(T+T_0)/Det$;
\begin{displaymath}
 Det=A e^{\alpha}-B e^{-\alpha}+C e^{\bar{\alpha}}-T e^{-\bar{\alpha}} ,
\end{displaymath}
\begin{displaymath}
 A=e^{\alpha} \bar{\alpha}^2 (r^2 e^{-2 \bar{\alpha}}-s^2 e^{2
   \bar{\alpha}})-2\alpha \bar{\alpha} e^{-\alpha} w+\alpha \bar{\alpha}
   e^{\alpha} z ( e^{-2\bar{\alpha}} r+ e^{2\bar{\alpha}} s),
\end{displaymath}
\begin{displaymath}
 A_0=e^{-\alpha} \bar{\alpha}^2 (r^2 e^{-2 \bar{\alpha}}-s^2 e^{2
   \bar{\alpha}})+2\alpha \bar{\alpha} e^{\alpha} z-\alpha \bar{\alpha}
   e^{-\alpha} w ( e^{2\bar{\alpha}} s+ e^{-2\bar{\alpha}} r),
\end{displaymath}
\begin{displaymath}
 B=e^{-\alpha} \bar{\alpha}^2 (r^2 e^{-2 \bar{\alpha}}-s^2 e^{2
   \bar{\alpha}})+2\alpha \bar{\alpha} e^{\alpha} z-\alpha \bar{\alpha}
   e^{-\alpha} w ( e^{-2\bar{\alpha}} r+ e^{2\bar{\alpha}} s),
\end{displaymath}
\begin{displaymath}
 B_0=e^{\alpha} \bar{\alpha}^2 (r^2 e^{-2 \bar{\alpha}}-s^2 e^{2
   \bar{\alpha}})-2\alpha \bar{\alpha} e^{-\alpha} w+\alpha \bar{\alpha}
   e^{\alpha} z ( e^{-2\bar{\alpha}} r+ e^{2\bar{\alpha}} s),
\end{displaymath}
\begin{displaymath}
 C=e^{-\alpha} \alpha \bar{\alpha} (w s e^{\bar{\alpha}-\alpha}-r z e^{
   \alpha-\bar{\alpha}})-\alpha e^{2\alpha+\bar{\alpha}} z(\alpha
   z-\bar{\alpha} s)+\alpha e^{-\alpha} w (\alpha
   e^{\bar{\alpha}-\alpha} w-
\bar{\alpha} e^{\alpha-\bar{\alpha}} r),
\end{displaymath}
\begin{displaymath}
 C_0=e^{\alpha} \alpha \bar{\alpha} (w s e^{\bar{\alpha}-\alpha}-r z e^{
   \alpha-\bar{\alpha}})-\alpha z (z\alpha e^{2\alpha-\bar{\alpha}}
   -\bar{\alpha} e^{\bar{\alpha}} s)+\alpha w (\alpha
   e^{-(\bar{\alpha}+2\alpha)} w-
\bar{\alpha} e^{-(2\alpha+\bar{\alpha})} r),
\end{displaymath}
\begin{displaymath}
 T=e^{-\alpha} \alpha \bar{\alpha} (z s e^{\bar{\alpha}+\alpha}-r w e^{-(
   \alpha+\bar{\alpha})})-\alpha \bar{\alpha} (e^{2\alpha-\bar{\alpha}}
   z r -e^{\bar{\alpha}} w s)+
\alpha^2 e^{-\alpha} (-e^{2\alpha} z^2+ e^{-2\alpha} w^2),
\end{displaymath}
\begin{displaymath}
 T_0=e^{\alpha} \alpha \bar{\alpha} (z s e^{\bar{\alpha}+\alpha}-r w e^{-(
   \alpha+\bar{\alpha})})-\alpha \bar{\alpha} (e^{-\bar{\alpha}}
   z r -e^{\bar{\alpha}-2\alpha} w s)+
\alpha^2 e^{\alpha} (-e^{2\alpha} z^2+ e^{-2\alpha} w^2),
\end{displaymath}
with $r= (1-\bar{\alpha} \mbox{Kn})$, $s=(1+\bar{\alpha} \mbox{Kn})$,
$w=(1-\alpha \mbox{Kn})$, $z=(1+\alpha \mbox{Kn})$. \newline To obtain a simple
solution which relates to the mean transport so long as only terms of
$O(\epsilon^2)$ are concerned, we see that if every term in the
x-momentum equation is averaged over an interval of time equal to
the period of oscillation, we obtain for our
solution as given by above equations the time-averaged (unit) body forcing
\begin{equation}
 \overline{\frac{\partial p}{\partial x}}=\epsilon^2 \overline{(\frac{\partial
 p}{\partial x})_2} =\epsilon^2 [\frac{D_{yyy}}{2 Re} + \frac{i Re}{4}
 (\phi \phi^*_{yy} -\phi^* \phi_{yy})] +O (\epsilon^3) =\epsilon^2
 \frac{\Pi_0}{Re} +O(\epsilon^3) ,   
\end{equation}
where $\Pi_0$ is the integration constant for the integration of
equation (12) and could be fixed indirectly in the coming equation
(22). Now, from Eqn. (14), we have
\begin{equation}
  D_y (\pm 1) \pm \mbox{Kn} D_{yy} (\pm 1)= -\frac{1}{2} [\phi_{yy} (\pm 1)+
  \phi^*_{yy} (\pm 1)] \mp \mbox{Kn}  \{\frac{1}{2} [\phi_{yyy} (\pm 1)
+\phi^*_{yyy} (\pm 1)] \} ,  
\end{equation}
where $D_y (y)= \Pi_0 y^2 +a_1 y+ a_2 + {\cal C} (y)$, and together
from equation (12), we obtain
\begin{displaymath}
 {\cal C} (y)=\frac{\alpha^2 Re^2}{2} [\frac{c_0 c_2^*}{g_1^2}
 e^{(\alpha+\bar{\alpha}^*) y} + \frac{c_0^* c_2}{g_2^2} e^{(\alpha+\bar{\alpha}) y} +
 \frac{c_0 c_3^*}{g_3^2} e^{(\alpha-\bar{\alpha}^*) y} + \frac{c_0^*
 c_3}{g_4^2} e^{(\alpha-\bar{\alpha}) y} +
\end{displaymath}
\begin{displaymath}
 \hspace*{3mm} \frac{c_1 c_2^*}{g_3^2} e^{ (\bar{\alpha}^*-\alpha) y} +
 \frac{c_1^* c_2}{g_4^2} e^{(\bar{\alpha}-\alpha) y}
 + \frac{c_1 c_3^*}{g_1^2} e^{-(\bar{\alpha}^*+\alpha) y}+ \frac{c_1^*
 c_3}{g_2^2} e^{-(\bar{\alpha}+\alpha) y} +
\end{displaymath}
\begin{equation}
 \hspace*{3mm} \frac{c_2 c_3^*}{g_5^2} e^{(\bar{\alpha}-\bar{\alpha}^*) y}+
 \frac{c_2^* c_3}{g_5^2} e^{(\bar{\alpha}^* -\bar{\alpha}) y} +
 2 \frac{c_2 c_2^*}{g_6^2}  e^{(\bar{\alpha}^*
 +\bar{\alpha}) y}+2 \frac{c_3 c_3^*}{g_6^2}  e^{-(\bar{\alpha}^*
 +\bar{\alpha}) y}]  ,
\end{equation}
with $g_1=\alpha+\bar{\alpha}^*$, $g_2=\alpha+\bar{\alpha}$,
$g_3=\alpha-\bar{\alpha}^*$, $g_4=\alpha-\bar{\alpha}$,
$g_5=\bar{\alpha}-\bar{\alpha}^*$,
$g_6=\bar{\alpha}+\bar{\alpha}^*$. In realistic applications we
must determine $\Pi_0$ from considerations of conditions at the ends
of the matter-region. $a_1$ equals to zero because of the symmetry of
boundary conditions.  \newline Once $\Pi_0$ is specified,
our solution for the mean speed ($u$ averaged over time) of matter-flow
is
\begin{equation}
 {U}=\epsilon^2 \frac{D_y}{2}=\frac{\epsilon^2}{2} \{ {\cal
 C}(y)-{\cal C} (1)+R_0-\mbox{Kn} \,{\cal C}_y (1)+\Pi_0 [y^2-(1+2 \mbox{Kn})] \} 
\end{equation}
where $R_0$ $=-\{ [\phi_{yy} (1)+\phi^*_{yy} (1)]$ $- \mbox{Kn}
[\phi_{yyy} (1)$ $+\phi^*_{yyy} (1)]\}/2$, which has a numerical
value about $3$ for a wide range of $\alpha$ and $Re$ (playing the role of viscous
dissipations) when Kn$=0$. To illustrate our results clearly, we adopt $U(Y) \equiv u(y$)
for the time-averaged results with $y\equiv Y$ in the following.
\section{Results and Discussion}
We check our approach firstly by examining $R_0$ with that of
no-slip (Kn$=0$) approach. This can be done easily once
we consider terms of $D_y(y)$ and ${\cal C} (y)$ because to
evaluate $R_0$ we shall at most take into account the higher
derivatives of $\phi(y)$, like $\phi_{yy} (y)$, $\phi_{yyy} (y)$
instead of $\phi_y (y)$ and escape from the prescribing of $a_2$.
\newline Our numerical calculations confirm that the mean streamwise
velocity distribution (averaged over time) due to the induced
motion by the wavy elastic vacuum-matter interface in the case of free (vacuum) pumping
is dominated by $R_0$
(or Kn) and the parabolic distribution $-\Pi_0 (1-y^2)$. $R_0$
which defines the boundary value of $D_y$ has its origin in the
y-gradient of the first-order streamwise velocity distribution, as
can be seen in Eqn. (14).
\newline In addition to the terms mentioned above, there is a
perturbation term which varies across the channel : ${\cal C} (y)
-{\cal C} (1)$. Let us define it to be
\begin{equation}
 F(y)=\frac{-200}{\alpha^2 Re^2} [{\cal C} (y) -{\cal C} (1)]
\end{equation}
To compare with no-slip (Kn$=0$) results, we plot
three cases, $\alpha=0.1, 0.4$, and $0.8$ for the same Reynolds number $Re=1$
 of our
results : Kn$=0.1$ with those Kn$=0$ into Fig. 1.
 We remind the readers that the Reynolds number
here is based on the wave speed. This figure
confirms our approaches since we can recover no-slip results
by checking curves of Kn$=0$ and finding them being almost
completely matched in-between. The physical trend herein is also
the same as those reported in Refs. [12-13] for the
slip-flow effects. The slip produces decoupling with the inertia of the wavy interface.
\newline Now, let us define a critical reflux condition as one for
which the mean velocity ${U} (Y)$ equals to zero at the
center-line $Y=0$ (cf. Fig. 1). With equations (12,20-21), we have
\begin{equation}
  \Pi_{0_{cr}}=Re \overline{(\frac{\partial p}{\partial x})_2}=\frac{[{\alpha^2
  Re^2}F(0)/200 +\mbox{Kn} \,{\cal C}' (1)-R_0]}{-(1+2 \mbox{Kn})}
\end{equation}
which means the critical reflux condition is reached when $\Pi_0$
has above value. Pumping against a positive (unit) body forcing
greater than the critical value would result in a backward
transport (reflux) in the central region of the stream. This
critical value depends on $\alpha$, $Re$, and Kn. There will be no
reflux if the (unit) body forcing or pressure gradient is smaller
than this $\Pi_0$. Thus, for some $\Pi_0$ values less than
$\Pi_{0_{cr}}$, the matter (flow) will keep moving or evolving
forward. On the contrary, parts of the matter (flow) will move or
evolve backward if $\Pi_0 > \Pi_{0_{cr}}$. This result could be
similar to that in [16] using different approach or qualitatively
related to that of [5] : even for very slow growth of $\Lambda$,
the gravitationally bound systems become unbound while the
nongravitationally bound systems remain bound for certain
parameters defined in [5] (e.g., $\eta$). \newline  We present
some of the values of $\Pi_0(\alpha, Re; \mbox{Kn}=0, 0.1)$
corresponding ot freezed or zero-volume-flow-rate states
($\int_{-1}^1 U(Y) dY=0$) in  Table 1 where the wave number
($\alpha$) has the range between $0.20$ and $0.80$; the Reynolds
number ($Re$)$=0.1,1,10,100$.  We observe that as Kn increases
from zero to 0.1, the critical $\Pi_0$ or time-averaged (unit)
body forcing decreases significantly. For the same Kn, once Re is
larger than 10, critical reflux values $\Pi_0$ drop rapidly and
the wave-modulation effect (due to $\alpha$) appears. The latter
observation might be interpreted as the strong coupling between
the vacuum-matter interface and the inertia of the streaming
matter-flow. The illustration of the velocity fields for those
zero-flux (zero-volume-flow-rate) or freezed states are shown in
Figure 2. There are three wave numbers : $\alpha=0.2, 0.5, 0.8$.
The Reynolds number is 10. Both no-slip and slip (Kn=0.1) cases
are presented. The arrows for slip cases are schematic and
represent the direction of positive and negative velocity fields.
\newline Some remarks could be made about these states : the
matter or universe being freezed in the time-averaged sense for
specific dissipations (in terms of Reynolds number which is the
ratio of wave-inertia and viscous effects) and wave numbers (due
to the wavy vacuum-interface or vacuum fluctuations) for either
no-slip and slip cases. This particular result might also be
related to a changing cosmological term (growing or decaying
slowly) or the critical density mentioned in Refs. [1-2]. If we
treat the (unit) body forcing as the pressure gradient, then for
the same transport direction (say, positive x-direction), the
negative pressure (either downsdtream or upstream) will, at least,
occur once the time-averaged flow (the maximum speed of the matter
(gas) appears at the center-line) is moving forward! For example,
for $\Pi_0=Re \overline{(\frac{\partial p}{\partial x})_2}=-10<0$
 ($Re=1, \alpha=0.5$), the velocity field (profile) is shown in Fig. 3.
One possible $p$-pair  for uniform (negative) gradient (mean value theorem): $p_{downstream}
-p_{upstream} <0$, with $x_{downstream}-x_{upstream}>0$ : $p_{downstream}<0$,
$p_{downstream}<p_{upstream} <0$.
\newline Meanwhile, the time-averaged transport induced by the wavy interface
is proportional to the square
of the amplitude ratio (although the small amplitude waves being presumed), as can be seen in Eqn. (12) or (20), which is qualitatively the
same as that presented in [9] for analogous interfacial problems.
In brief summary,
the entrained transport (pattern, either postive or negative and there is possibility : freezing)
due to the wavy vacuum-matter interface
 is mainly tuned
by the (unit) body forcing or $\Pi_0$ for fixed Re. Meanwhile,
$\Pi_{0_{cr}}$ depends strongly on the Knudsen number (Kn, a
rarefaction measure) instead of Re or $\alpha$. We hope that in
the future we can investigate other issues like the role of phase
transition and that of cyclic universes [16-18] using the present
or more advanced approach.

\newpage
%

\vspace{2mm}

\psfig{file=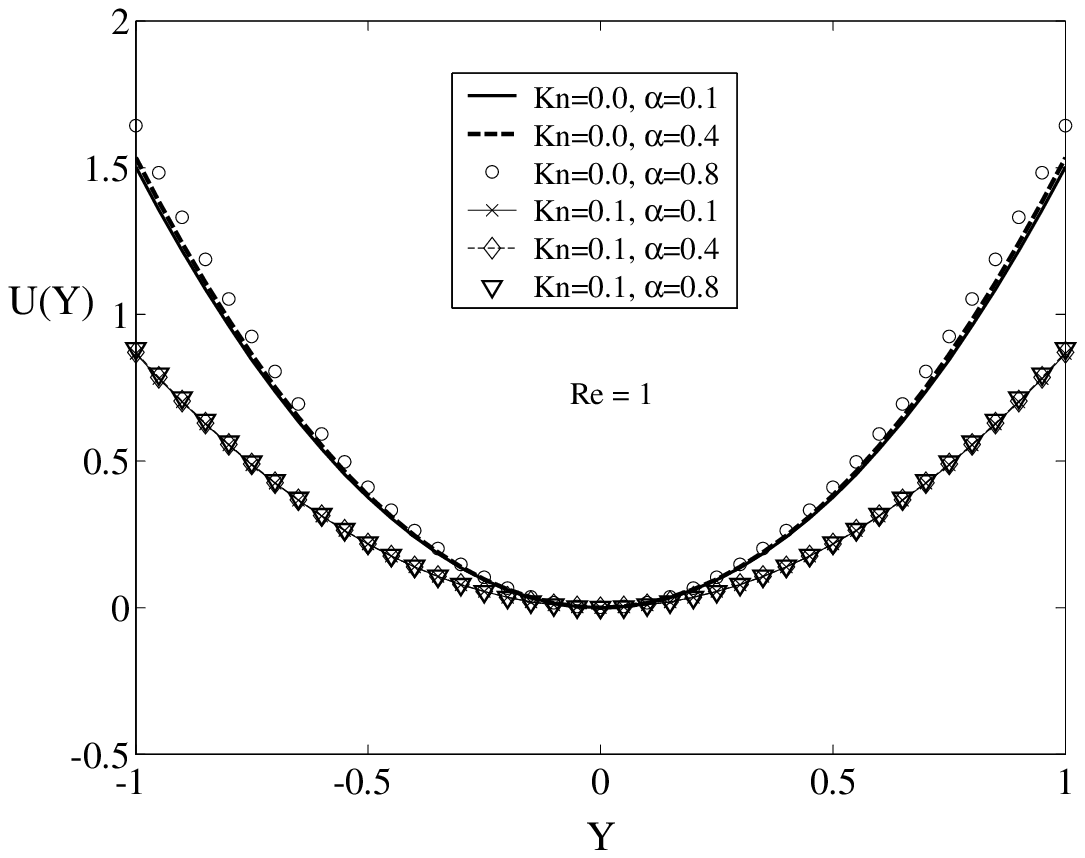,bbllx=0cm,bblly=13.8cm,bburx=12cm,bbury=24cm,rheight=10cm,rwidth=10cm,clip=}
%
\begin{figure} [h]

\hspace*{3mm} Fig. 1 \hspace*{1mm} Demonstration of Kn and
$\alpha$ effects on the time-averaged velocity (fields) profiles.
\newline \hspace*{3mm} $\Pi_0 = {\Pi_0}_{cr}$. The Reynolds number
(the ratio of the wave-inertia and viscous dissipation \newline
\hspace*{3mm} effects) is 1. $U(Y)=0$ at $Y=0$. Kn is the Knudsen
number (a rarefaction measure).
\end{figure}

\vspace{3mm}
\psfig{file=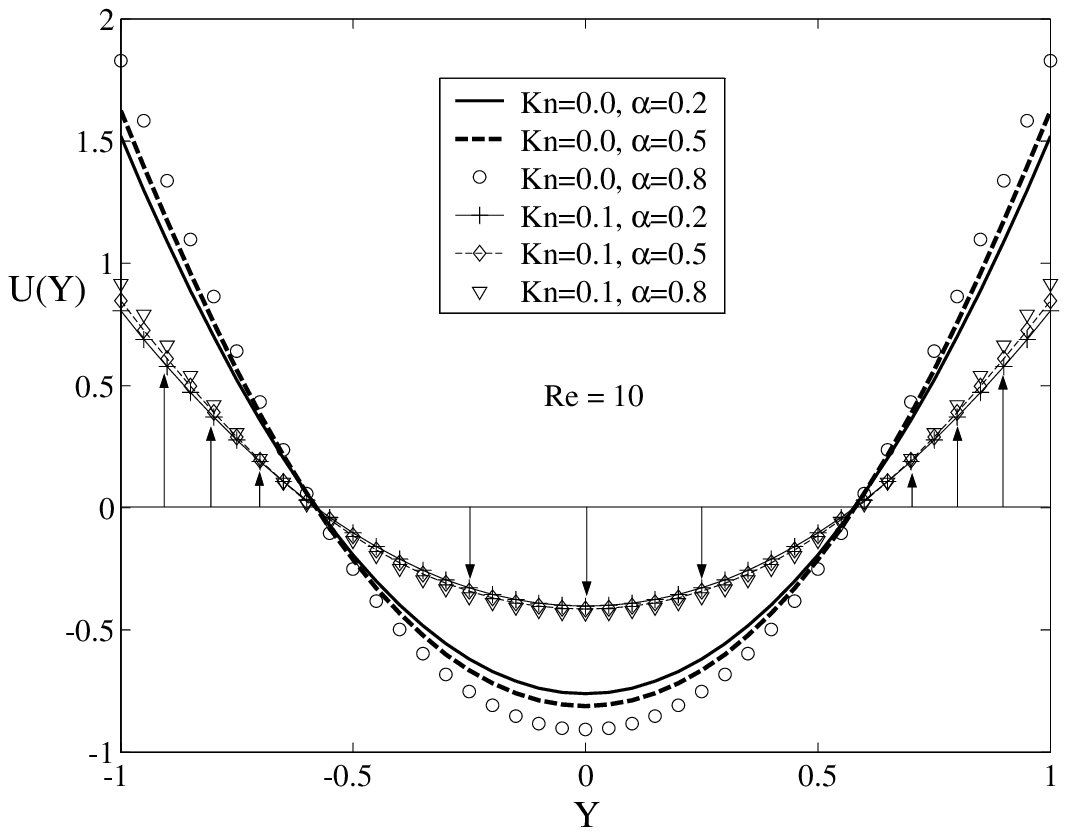,bbllx=0.0cm,bblly=12.5cm,bburx=12cm,bbury=24cm,rheight=9.8cm,rwidth=9.8cm,clip=}
%
\begin{figure} [h]
\hspace*{3mm} Fig. 2 \hspace*{1mm} Demonstration of the zero-flux
states : the mean velocity field $U(Y)$ for \newline \hspace*{3mm}
wave numbers $\alpha=0.2,0.5,0.8$. The Reynolds number is $10$. Kn
is the rarefaction measure
\newline \hspace*{3mm}
(the mean free path of the particles divided by the characteristic length).
\newline \hspace*{3mm}
The arrows are schematic and illustrate the directions of positive and negative $U(Y)$.
\newline \hspace*{3mm}
The integration of $U(Y)$ w.r.t. $Y$ for these velocity fields gives zero volume flow rate.
\end{figure}

\newpage

\psfig{file=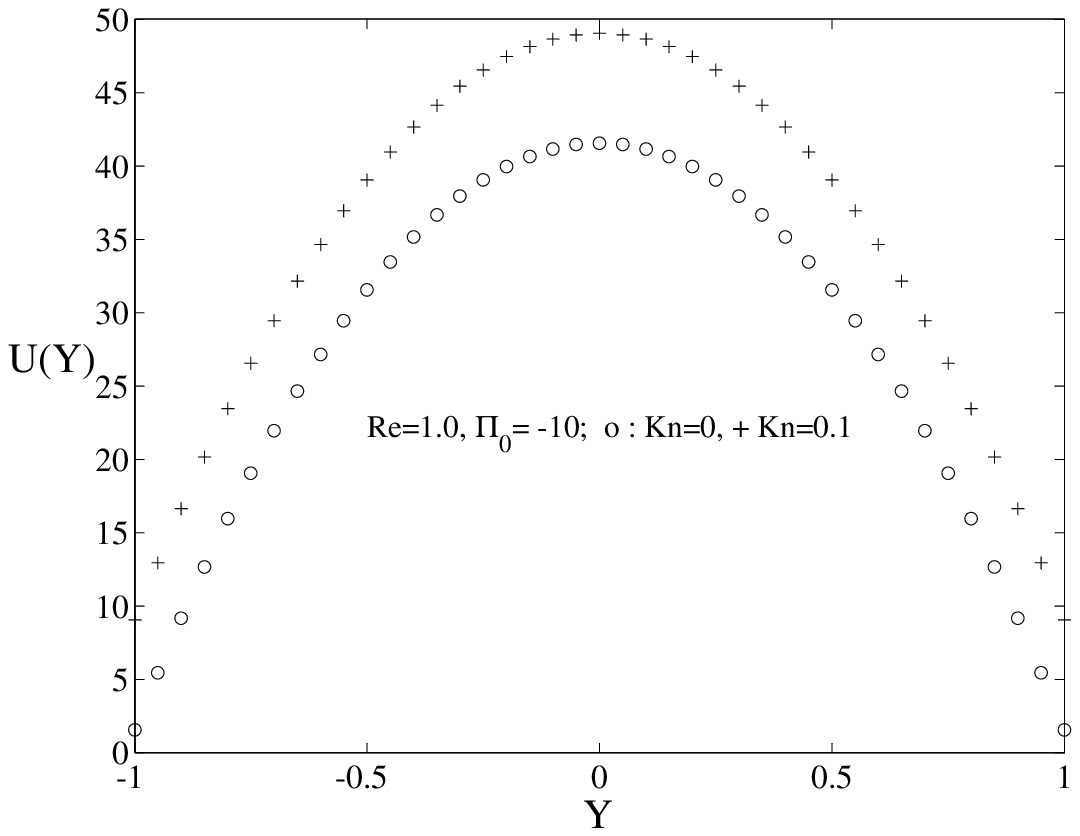,bbllx=0.0cm,bblly=12.5cm,bburx=12cm,bbury=24cm,rheight=9.8cm,rwidth=9.8cm,clip=}

\begin{figure}  [h]
\hspace*{3mm} Fig. 3 \hspace*{1mm}Demonstration of the
negative-$p$ states : the mean velocity field $U(Y)$ \newline
\hspace*{3mm} for the wave number $\alpha=0.5$ and the Reynolds
number $Re=1$.
\end{figure}

\newpage

\begin{table}[h]
 \caption{Zero-flux or freezed states values ($\Pi_0$) for a flat vacuum-matter interface.}
\vspace*{5mm}
\begin{center}
\begin{tabular}[b]{|r|c|c|c|c|c|}      \hline
    &          &  Re  &   &          &      \\ \hline
 Kn & $\alpha$ &  0.1 & 1 & 10 &   100 \\ \hline
 0  
    & 0.2  &  4.5269  & 4.5269  &   4.5231       &              4.3275  \\  \cline{2-6}
    & 0.5  &  4.6586  & 4.6584  &   4.6359       &              4.2682  \\  \cline{2-6}
    & 0.8  &   4.9238 &  4.9234 &    4.8708      &               4.4488  \\ \hline 
0.1 
     & 0.2 &  2.4003  & 2.4000  &   2.3774       &              1.2217    \\ \cline{2-6}
     & 0.5 &  2.4149  & 2.4132  &   2.2731       &             -0.9054 \\ \cline{2-6}
     & 0.8 &  2.4422  & 2.4379  &   2.0718       &             -3.4151  \\ \hline 
\end{tabular}            
\end{center}
\end{table}
\end{document}